# Bayesian algorithms for recovering structure from single-particle diffraction snapshots of unknown orientation: a comparison

Brian Moths and Abbas Ourmazd[1]
Dept. of Physics, University of Wisconsin Milwaukee, 1900 E. Kenwood Blvd, Milwaukee, WI 53211, USA

## Abstract

The advent of X-ray Free Electron Lasers promises the possibility to determine the structure of individual particles such as microcrystallites, viruses and biomolecules from single-shot diffraction snapshots obtained before the particle is destroyed by the intense femtosecond pulse. This program requires the ability to determine the orientation of the particle giving rise to each snapshot at signal levels as low as $\sim 10^{-2}$ photons/pixel. Two apparently different approaches have recently demonstrated this capability. Here we show they represent different implementations of the same fundamental approach, and identify the primary factors limiting their performance.

---

[1] Corresponding author (Ourmazd@uwm.edu)

## 1. Introduction

X-ray Free Electron Lasers promise to move crystallography beyond crystals. For example, moves are afoot to determine the structure of biological molecules and their assemblies by exposing a succession of individual single particles to intense femtosecond pulses of X-rays (Solem & Baldwin, 1982, Neutze *et al.*, 2004, Gaffney & Chapman, 2007). In addition to experimental issues, two algorithmic challenges must be overcome in order to recover structure from such diffraction snapshots. First, the orientation of the object giving rise to each snapshot must be determined. Second, this must be performed at extremely low signal. A typical 500kD biomolecule, for example, scatters only 100 of the $\sim 10^{12}$ incident photons, with the photon count per pixel being as low as $10^{-2}$ at the detector (Shneerson *et al.*, 2008). As the particle orientations giving rise to the snapshots are unknown, signal cannot be boosted by averaging, and orientation recovery must be carried out at "raw signal level" in the presence of shot (Poisson) and background scattering noise (Shneerson *et al.*, 2008, Fung *et al.*, 2009). Orientation recovery is thus one of the most critical steps in single-particle structure determination (Leschziner & Nogales, 2007). Once diffraction pattern orientations have been discovered, the 3-dimensional (3D) diffraction volume can be assembled and the particle structure recovered by standard phasing algorithms (Gerchberg & Saxton, 1972, Fienup, 1978, Miao *et al.*, 2001, Shneerson *et al.*, 2008, Fung *et al.*, 2009, Loh & Elser, 2009).

Using an adaptation of Generative Topographic Mapping (GTM) (Bishop *et al.*, 1998), Fung et al. (Fung *et al.*, 2009) published the first successful recovery of the structure of a molecule from simulated diffraction snapshots of unknown orientation at signal levels expected from a 500kD molecule by utilizing the information content of the entire ensemble of diffraction snapshots. Loh and Elser (Loh & Elser, 2009) demonstrated structure recovery from simulated diffraction snapshots by an apparently different approach, using a so-called Expansion-Maximization-Compression (EMC) algorithm (Loh & Elser, 2009). Here, we show these two approaches are fundamentally the same, and discuss their capabilities and limitations.

In order to facilitate the discussion, the structure recovery process is divided into two steps: a) orienting the diffraction snapshots and assembling the 3D diffraction volume; and b) recovering the structure by a phasing algorithm. Since we are concerned with orientation recovery, the discussion will be confined to the first step.

The differences in presentation and notation notwithstanding, the two approaches are the same in all essential features. Specifically, they both:
**a.** Exploit the information content of the entire dataset;
**b.** Recognize that a nonlinear mapping function relates the space of object orientations to the space of scattered intensities;
**c.** Determine the mapping function by Bayesian inference;
**d.** Use the well-established expectation-maximization (EM) iterative algorithm (Dempster *et al.*, 1977) to maximize likelihood;
**e.** Apply a constraint to guide likelihood maximization; and
**f.** Implement noise-robust algorithms with essentially the same computational scaling behaviors.

The primary difference between the two approaches concerns the way the step (e) is implemented. This paper elucidates the essential similarity between these two approaches, thus clarifying the basis of Bayesian approaches to orienting snapshots. Details of each approach can be found in the cited references. To facilitate a detailed comparison of the two papers, Appendix 1 provides a translation table for the symbols used in each.

**2. Conceptual outline of orientation recovery**

In essence, diffraction from a given object is a process ("a machine"), which takes an orientation as input to generate a diffraction pattern as output. With a detector consisting of $p$ pixels and the pixel intensities as coordinates, one can represent a diffraction pattern as a point in a $p$-dimensional Euclidean space of intensities. The information content of each diffraction pattern can be captured by ensuring that the pixels represent Shannon-Nyquist samples. In this picture, diffraction maps an orientation to a point in a $p$-dimensional space. Because an object has only three orientational degrees of freedom ("Euler angles"), in the absence of noise, the points in the $p$-dimensional space of intensities define a 3D manifold, which is, in fact, a nonlinear map of the SO(3) manifold of orientations[2].

The representation of object orientations bears careful consideration. Despite their wide-spread use, Euler angles are not a good representation of orientational similarity, because an object can be rotated through large Euler angles $(\alpha, \beta, \gamma)$ and end at an orientation very close to its starting point. As the Euclidean distance in quaternion space is a good measure of (dis)similarity between orientations, both Fung and LE use unit quaternions (Kuipers, 2002) to represent orientations. Diffraction, therefore, can be thought of as a (real) function $y(x)$, with $x$ representing a unit quaternion.

A diffraction snapshot consists of $p$ intensity values. The mapping thus takes an orientation $x$ to generate a model snapshot $\mathbf{y}(x) = (y_1, \ldots, y_p)$. These are to be compared with experimental snapshots $\mathbf{t} = (i_1, \ldots, i_p)$, but will, in general, not be identical to any single snapshot due to (experimental) noise[3].

Because a given object has only three orientational degrees of freedom, the points $\mathbf{t}_n = (i_1, \ldots, i_p)$ representing the diffraction snapshots in the so-called manifest intensity space trace out a 3D manifold, which is a nonlinear map of the SO(3) manifold of orientations. Given the "input" and "output" manifolds, it is conceptually straightforward to discover the nonlinear map $y$ linking the two, and thus assign an orientation to each diffraction snapshot. Once this has been accomplished, snapshots of similar orientation can be averaged to boost signal, and structure recovery can proceed by standard techniques. In fact, appropriately wielded, manifold embedding can improve signal far more efficiently than simple averaging of similar snapshots (Schwander *et al.*, 2010), but this is beyond the scope of the present paper.

---

[2] In mathematical terms, diffraction is a mapping $\mathcal{M} := \Phi[SO(3)] \subset L^2(\mathbb{R}^2)$, with $\Phi$ describing the diffraction process. In the absence of object symmetry, the map is one-to-one and onto.

[3] In this paper, vectors are represented by bold lower-case, matrices by bold upper-case letters.

We now discuss how this conceptual outline is implemented in the two apparently different approaches by Fung et al. (Fung *et al.*, 2009) (hereafter Fung) and Loh and Elser (Loh & Elser, 2009) (hereafter LE).

## 3. Exploiting the information content of the dataset

Both Fung and LE use the conceptual framework that snapshot orientations can be determined by discovering the nonlinear map connecting the two manifolds. The power of this approach stems from the fact that the intensity manifold is defined by the entire ensemble of snapshots. In essence, one is using the whole dataset to assign an orientation to each snapshot. This is needed to overcome the paucity of information in any single snapshot. Key here is the recognition that the "mutual information" between the snapshots of a large ensemble is much larger than the information in any single snapshot (Fung *et al.*, 2009, Elser, 2009).

To render the formalism tractable, the SO(3) space of orientations is represented by a discrete set of $K$ orientations ("nodes") $\{x_k\}$, distributed nearly uniformly on the three-sphere (Lovisolo & da Silva, 2001, Coxeter, 1973). The inter-node spacing is chosen to satisfy the Shannon-Nyquist sampling criterion, determined as follows. Consider recovering the structure of an object with largest diameter $D$ (radius $R$) to resolution $r$ (Fig. 1). The orientational accuracy needed is then:

$$\Delta\theta_{shannon}^{orient} = \frac{1}{2}\frac{r}{R} = \frac{r}{D} \quad , \tag{1}$$

with the number of independent orientations in 3D given by:

$$N_{nodes} = \frac{1}{2} \cdot \left( \frac{\text{Area of 3-sphere} = 2\pi^2}{(\text{Shannon element on 3-sphere})^3} \right) \cdot \frac{1}{\text{\# of symmetry elements}} \quad . \tag{2}$$

The pre-factor of ½ accounts for the fact that the 3-sphere is a double-cover of SO(3). The Shannon element in terms of quaternions $q$ is:

$$\Delta q_{shannon} = \sqrt{2\left[1-\cos\frac{\Delta\theta_{shannon}}{2}\right]} \approx \frac{\Delta\theta_{shannon}}{2} \quad , \tag{3}$$

leading to:

$$N_{nodes} = \frac{8\pi^2}{S(\Delta\theta_{shannon})^3} \quad , \tag{4}$$

where S is the number of symmetry elements of the molecule being reconstructed.

The information content of the dataset is compromised by noise. Noise is handled by Fung et al. via a Gaussian model for the departures of a vector representing a noisy snapshot from its ideal noise-free position in the *p*-dimensional intensity space. The large number of pixels used as components of a vector representing a snapshot ensures, via the Central Limit Theorem, that a Gaussian model is appropriate regardless of the specific noise spectrum present in each pixel.

This approach can thus deal with substantial background scattering (Schwander *et al.*, 2010). LE uses a Poisson noise model. As pointed out by LE, it remains to be established whether this is sufficient to deal with situations where other types of noise also play a role (Loh & Elser, 2009).

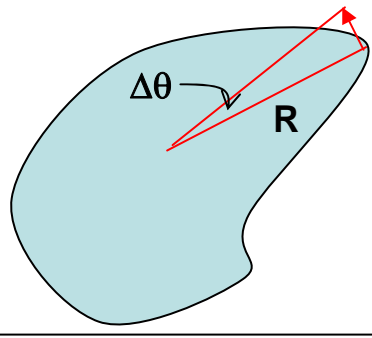

**Fig. 1.** Schematic relationship between object diameter D (=2R), spatial resolution r, and required orientational accuracy $\Delta\theta$

4. **Bayesian inference and likelihood maximization**

To link the orientations $\{x_k\}$ to intensity space, both approaches use Bayesian inference and iterative likelihood maximization. Given a pair of events A and B with marginal probabilities $P(A)$ and $P(B)$, Bayes' theorem links their conditional probabilities via the expression:

$$P(A|B) = \frac{P(B|A)P(A)}{P(B)} \quad . \tag{5}$$

This is used to link the space of orientations with the space of observed diffraction snapshots. Starting with an initial guess for the nonlinear map, the likelihood of the observed data, given the model snapshots $\{\mathbf{y}(x_k)\}$ is:

$$L = \prod_{n=1}^{N} \sum_{k=1}^{K} p(\mathbf{t}_n | \mathbf{y}(x_k)) p(x_k) \quad , \tag{6}$$

where $\mathbf{t}_n$ and $\mathbf{y}(x_k)$ represent the actual and model snapshots, and the indices *n* and *k* run over the set of *N* diffraction patterns and *K* orientations, respectively. The probability $p(\mathbf{t}_n | \mathbf{y}(x_k))$ is determined by the noise model, and $p(x_k)$ is the prior probability of the orientation $x_k$, which is $1/K$ when all orientations are equally likely.

Both Fung and LE maximize the log-likelihood iteratively by the well-known Expectation-Maximization (EM) algorithm (Dempster *et al.*, 1977). Each iteration modifies the model snapshots, effectively moving the manifold defined by them closer to the experimental data. There is no guarantee that the final solution is a global maximum.

Once the mapping corresponding to maximum likelihood has been determined, the orientation of each measured diffraction pattern $\mathbf{t}_n$ is taken to be that $x_k$ which maximize the probability of $\mathbf{t}_n$ "belonging" to the nearest model diffraction patter $\mathbf{y}(x_k)$ given by the value of *k* corresponding to the maximum of $p(x_k | \mathbf{t}_n)$, viz. $k = \arg\max_j p(x_j | \mathbf{t}_n)$.
The conditional probability $p(x_k | \mathbf{t}_n)$ is determined using Eq. (5).

Having assigned the *N* diffraction snapshots to the *K* orientational bins, the diffraction volume can be reconstructed. In standard "classification and averaging," diffraction patterns assigned to the same orientation $x_k$ are averaged so that there is one representative diffraction pattern for each $x_k$. So-called generative models such as that used by Fung allow one to construct ("generate") model snapshots for each orientation directly from the manifold. As the manifold represents the information content of the entire dataset, the generative approach offers significantly greater noise reduction than classification and averaging, which relies on the information in the neighborhood of a given orientation only (Schwander *et al.*, 2010).

Each averaged or generated snapshot is placed in reciprocal space according to its orientation, resulting in a set of irregularly spaced points in reciprocal space. These are interpolated onto a Cartesian grid so as to allow fast Fourier transformation during iterative phasing (Gerchberg & Saxton, 1972, Schwander *et al.*, 2010, Fienup, 1978).

5. **Constraints to guide expectation-maximization**
The only substantive difference between the GTM and the EMC algorithms is the way in which the manifold embedding process is implemented, more specifically, the way the model diffraction patterns are evolved so as to maximize the likelihood. In principle, one would modify the model diffraction patterns along steepest ascent in log-likelihood, until the derivative with respect to changes in the model diffraction patterns is zero. However, this approach is too simple to be of use in practice. Suppose we have found the map *y* such that *L* is maximized, and suppose we now exchange a pair of model images assigned to $x_1$ and $x_2$, viz. $y(x_1) \rightleftarrows y(x_2)$. This simply switches the order of the first two terms in the sum over *k* in Eq. (6), leaving the likelihood unchanged. By the same reasoning, we are able to permute the images assigned to the $x_k$ arbitrarily without changing the likelihood *L*. This means that likelihood maximization alone is unable to find a unique solution, and is, for example, unable to distinguish between the two very different neighborhood assignments shown in Fig. 2.

In order to eliminate this problem, both the GTM and EMC algorithms implement a "contiguity constraint" on the map *y*. This constraint demands that two nodes which are close to each other in the space of orientations be mapped to points close to each other in data space. Fung and LE implement this contiguity constraint differently. In the GTM approach used by Fung, the map is expanded in terms of a set of basis functions:

$$\mathbf{y}(x) = \sum_{m=1}^{M} \varphi_m(x) \mathbf{w}_m , \qquad (7)$$

where $\varphi_m$ is one of *M* basis functions ( *M* < the number of independent orientations *K*), and $\mathbf{w}_m$ represent the expansion coefficients (weights). Likelihood maximization proceeds by adjusting the *M* sets of *p* coefficients. The basis functions are chosen so as to vary slowly with $x$. In the current implementation of GTM, they are Gaussians (Bishop *et al.*, 1998). The map in Eq (7) varies slowly, provided the weights $\mathbf{w}_m$ are small. This is achieved by imposing a zero-centered Gaussian distribution on the sum of the squares of the weights. This strategy helps ensure that, topologically, the neighborhood assignments in manifest space reflect the neighborhood assignments in latent (orientation) space, i.e, $\mathbf{y}(x_k)$ is close to $\mathbf{y}(x_{k'})$ when $x_k$ is close to $x_{k'}$.

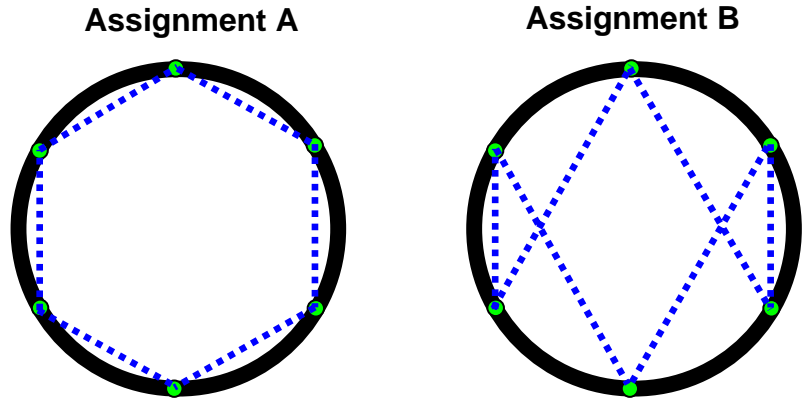

**Fig. 2.** The two different neighborhood assignments indicated by the dotted lines have the same likelihood. Assignment A is clearly preferred to Assignment B, which increases the distances between adjacent neighbors. An additional "contiguity constraint" is required to distinguish between these two assignments. The black circle represents the "true" data manifold, the green dots the model images $\mathbf{y}(x_k)$, the dotted blue lines the neighborhood assignments ("fitted" manifold).

The EMC algorithm of LE, in contrast, uses the model diffraction patterns $\mathbf{y}(x_k)$ themselves (rather than the weights $\mathbf{w}_m$) as fitting parameters. After each expectation-maximization step, a so-called "compression" step inserts the model diffraction patterns $\mathbf{y}(x_k)$ into reciprocal space according to their orientations, and the resulting irregularly spaced points are interpolated onto a uniform grid to determine a new diffraction volume by local averaging. Next, an "expansion" step uses the new diffraction volume as the source for a fresh set of model diffraction snapshots by interpolating back onto the irregularly spaced points corresponding to the pixels of each of the model diffraction patterns. In this approach, both the compression and expansion steps act as low pass filters; replacing two diffraction patterns by their average and then deducing two diffraction patterns from the average removes sharp variations between diffraction patterns mapped to similar points in reciprocal space. In essence, the so-called compression-expansion cycle is an alternative implementation of the contiguity constraint, whereby neighboring orientations in latent space give rise to neighboring points in manifest intensity space.

The apparently different implementations of the contiguity constraint described above belie the fundamental similarity of the two approaches even in this step. As shown in Appendix 2, the Fung and LE approaches reduce to the same algorithm if the so-called weight regularization parameter is set to zero in Fung and the Compression-Expansion step is removed from the LE approach.

**6. Scaling behavior**

The fundamental similarities between the two approaches result in similar scaling in computational behavior. In brief, the computational demands rise as $E^n$, where $E = \left(\dfrac{D}{r}\right)^s$ is the number of resolution elements, $D$ and $r$ the object diameter and spatial resolution, and $s$ the number of orientational degrees of freedom. Typically, $2 \leq n \leq 3$, i.e., the computational cost scales as the sixth to ninth power of $\left(\dfrac{D}{r}\right)$ (Fung *et al.*, 2009), severely limiting the achievable resolution, and/or amenable object size. Significant improvements in this behavior are essential, with the most obvious route involving more efficient implementation and parallelization (Fung *et al.*, 2009, Loh & Elser, 2009). Fundamentally, however, the high computational cost of Bayesian approaches stems from their generality. It has been long known that the most general algorithms are the most inefficient and the way to improve this involves introducing problem-specific constraints (Le Cun *et al.*, 1990, Schwander *et al.*, 2010). This is the basis of a new generation of algorithms, which directly incorporate the physics of scattering.

**7. Summary and conclusions**

Bayesian approaches are currently the only published methods capable of orienting snapshots containing as few as 100 scattered photons and ~$10^{-2}$ per pixel. The present paper establishes that two apparently different Bayesian approaches to orienting diffraction snapshots are the same in all essential features. The elucidation of these features can guide the development of computationally more efficient algorithms, which are needed if the large and more complex datasets anticipated from ongoing experiments are to be successfully analyzed. The remarkable capability of the Fung and LE approaches to operate at extremely low signals stems not from algorithmic details, but from the realization that much of the information about a given snapshot resides not in the snapshot itself, but in the other snapshots in the dataset, and the entire information content is needed to orient each snapshot at low signal. This feature must form the basis of more advanced algorithms.

We acknowledge valuable discussions with Veit Elser, Russell Fung, Dilano Saldin, Peter Schwander, Pierre Thibault, and Chun Hong Yoon. This work was partially supported by award DE-SC0002164 from the Office of Basic Energy Sciences of the US Department of Energy.

# Appendix 1: Indices and symbols

Translation tables for indices and symbols used in Fung (Fung *et al.*, 2009) and LE (Loh & Elser, 2009)

**Indices**

| Fung | LE | Description |
|------|-----|-------------|
| $k$ | $j$ | Indexes the set of orientations corresponding to the model diffraction patterns |
| $d$ | $i$ | Indexes the pixels in an experimental or model diffraction pattern |
| $n$ | $k$ | Indexes the set of experimental diffraction patterns |

**Symbols**

| Fung | LE | Description |
|------|-----|-------------|
| **T** | **K** | Matrix whose entries are the pixel intensities of the experimental diffraction patterns. |
| **Y** | **W** | Matrix whose entries are the pixel intensities of the model diffraction patterns. |
| **R** | **P** | Matrix whose entries are the conditional probabilities of the model diffraction patterns, given the experimental diffraction patterns, e.g., $R_{kn}$ is the probability of the $k^{\text{th}}$ model diffraction pattern, given the $n^{\text{th}}$ experimental diffraction pattern. |

# Appendix 2: Comparison between contiguity constraint implementations

For GTM, the equation obtained from setting the derivative of the likelihood with respect to the model parameters is:

$$(\mathbf{\Phi}^{\text{T}}\mathbf{G}\mathbf{\Phi} + \lambda\mathbf{I})\mathbf{W} = \mathbf{\Phi}^{\text{T}}\mathbf{RT} \quad , \qquad (1)$$

where the matrix **G** is $K \times K$ diagonal matrix with entries given by $g_{kk} = \sum_n r_{kn}$.

In LE, the model parameters are the pixel intensities themselves, so **Φ** is the identity matrix and **W**=**Y**. There is no weight regularization in the EMC algorithm, i.e., $\lambda=0$. Therefore, Eq. (1) reduces to:

$$\mathbf{GY} = \mathbf{RT} \quad . \qquad (2)$$

This is to be compared with Eq. (11) of LE, which, translated into the same notation as Eq. (2) above, becomes $y_{kd} = \dfrac{\sum_n r_{kn} t_{nd}}{\sum_n r_{kn}}$. From the definition of the matrix **G**, it is clear that the LE update rule is given by $\mathbf{Y} = \mathbf{G}^{-1}\mathbf{RT}$, which is equivalent to Eq. (2) above.

**References**

Bishop, C. M., Svensen, M. & Williams, C. K. I. (1998). *Neural Computation* **10**, 215-234.

Coxeter, H. S. M. (1973). *Regular Polytopes*. New York: Dover.

Dempster, A. P., Laird, N. M. & Rubin, D. B. (1977). *Journal of the Royal Statistical Society. Series B (Methodological)* **39**, 1-38.

Elser, V. (2009). *IEEE Trans Information Theory* **55**, 4715 - 4722

Fienup, J. R. (1978). *Optics Letters* **3**, 27-29.

Fung, R., Shneerson, V., Saldin, D. K. & Ourmazd, A. (2009). *Nat Phys* **5**, 64-67.

Gaffney, K. J. & Chapman, H. N. (2007). *Science* **316**, 1444-1448.

Gerchberg, R. W. & Saxton, W. O. (1972). *Optik* **35**, 237–246.

Kuipers, J. B. (2002). *Quaternions and rotation sequences* 5ed. Princeton, N.J. : Princeton University Press, ©1999.

Le Cun, Y., Denker, J. S., Solla, S. A., Jackel, L. D. & Howard, R. E. (1990). Optimal brain damage, Advances in neural information processing systems 2, pp. 598-605. Denver: Morgan Kaufmann Publishers Inc.

Leschziner, A. E. & Nogales, E. (2007). *Annu. Rev. Biophys. Biomol. Struct.* **36**, 20.

Loh, N.-T. D. & Elser, V. (2009). *Phys. Rev. E* **80**, 026705.

Lovisolo, L. & da Silva, E. A. B. (2001). *IEE Proceedings - Vision, Image, and Signal Processing* **148**, 187-193.

Miao, J., Hodgson, K. O. & Sayre, D. (2001). *Proc Natl Acad Sci U S A* **98**, 6641-6645.

Neutze, R., Huldt, G., Hajdu, J. & van der Spoel, D. (2004). *Radiation Physics and Chemistry* **71**, 905-916.

Schwander, P., Fung, R., Phillips, G. N. & Ourmazd, A. (2010). *New J. Phys.* **12**, 1-15.

Shneerson, V. L., Ourmazd, A. & Saldin, D. K. (2008). *Acta Cryst. A* **64**, 303-315.

Solem, J. C. & Baldwin, G. C. (1982). *Science* **218**, 229-235.